\documentclass[%
 preprint, 
 amsmath,amssymb,
 aps, physrev,
]{revtex4-2}
\usepackage{amsmath,amssymb}
\usepackage{graphicx}
\usepackage{amsfonts}
\usepackage{mathrsfs}
\usepackage{xcolor}
\usepackage{mdframed}
\usepackage{booktabs}
\usepackage{braket}
\usepackage{bm}
\usepackage{algorithm}
\usepackage{algorithmicx}
\usepackage{algpseudocode}
\usepackage{listings}
\usepackage{hyperref}   

\newcommand{\ib}{\mathbf{i}}

\newcommand{\bx}{\bm{x}}

\newcommand{\Ab}{\mathbf{A}}

\newcommand{\Gb}{\mathbf{G}}

\newcommand{\Mb}{\mathbf{M}}

\newcommand{\Qb}{\mathbf{Q}}

\newcommand{\Ub}{\mathbf{U}}

\newcommand{\bfi}{\bm{i}}

\def\Pr{\mathop{\rm Pr}\nolimits}
\newcommand{\bfsym}[1]{\ensuremath{\boldsymbol{#1}}}
\def\bsX{\bfsym X}

\newcommand{\cH}{\mathcal{H}}

\newcommand{\cO}{\mathcal{O}}
\newcommand{\cP}{\mathcal{P}}

\newcommand{\lt}{\left}
\newcommand{\rt}{\right}

\newcommand{\PP}{\mathbb{P}}
\newcommand{\RR}{\mathbb{R}}

\def \bx{\mathbf{x}}

\def \max{\mbox{max}}

\def \Var{\mbox{Var}}

\newmdenv[
  backgroundcolor=blue!5,
  linecolor=addblue,
  linewidth=1.2pt,
  topline=true, bottomline=true,
  leftline=true, rightline=false,
  innerleftmargin=6pt, innerrightmargin=6pt,
  innertopmargin=4pt, innerbottommargin=4pt,
  skipabove=6pt, skipbelow=6pt
]{addedframe}

  {\hfill$\square$\par\medskip}

\newtheorem{theorem}{Theorem}[section]
\newtheorem{lemma}[theorem]{Lemma}
\newtheorem{remarka}[theorem]{Remark}
\newenvironment{remark}{\begin{remarka} \rm}{\end{remarka}}

\newtheorem{corollary}[theorem]{Corollary}

\begin{document}

\title{Quantum Statistical Bootstrap}

\author{Yongkai Chen}
\affiliation{Department of Statistics, University of Georgia, Athens, GA 30602, USA}
\affiliation{Department of Statistics, Harvard University, Cambridge, MA 02138, USA}

\author{Ping Ma}
\email{pingma@uga.edu}
\affiliation{Department of Statistics, University of Georgia, Athens, GA 30602, USA}

\author{Wenxuan Zhong}
\email{wenxuan@uga.edu}
\affiliation{Department of Statistics, University of Georgia, Athens, GA 30602, USA}

\date{\today}

\begin{abstract}
The bootstrap is a foundational tool in statistical inference, but its classical implementation relies on Monte Carlo resampling, introducing approximation error and incurring high computational cost---especially for large datasets and complex models. We present the Quantum Bootstrap (QBOOT), a quantum algorithm that computes the ideal bootstrap estimate exactly by encoding all possible resamples in quantum superposition, evaluating the target statistic in parallel, and extracting the aggregate via quantum amplitude estimation. Under mild circuit efficiency assumptions, QBOOT achieves a near-quadratic speedup over the classical bootstrap in approximating the ideal estimator, independent of the statistic or underlying distribution. We provide a rigorous theoretical analysis of its statistical error properties---addressing a gap in the quantum algorithms literature---and validate our results through experiments on the IBM quantum simulator for the sample mean problem. Our findings demonstrate that QBOOT preserves the asymptotic properties of the ideal bootstrap while substantially improving computational efficiency and precision, establishing a scalable and principled framework for quantum statistical inference.
\end{abstract}

\maketitle


\section{Introduction}

Recent advances in quantum computing have made general‑purpose, programmable quantum devices—such as IBM Quantum Experience, Microsoft Quantum, and Amazon Braket—publicly accessible. These quantum devices, often termed \emph{noisy intermediate‑scale quantum} (NISQ) systems \citep{preskill2018quantum}, are characterised by noise and limited computing unit/qubit counts, although some now reach the scale of thousands of qubits \citep{castelvecchi2023ibm}. Quantum computing has already demonstrated quantum advantage on carefully designed, highly physics‑oriented tasks \citep{arute2019quantum,zhong2020quantum,wu2021strong,Acharya2025}. Given that many statistical procedures are computationally demanding, these developments have naturally prompted the questions: \emph{Can quantum computers address core bottlenecks in statistics? If so, which problems are best‑suited to quantum speedups?} Despite recent progress, a clear answer remains elusive. See, e.g., \citet{wang2022quantum,ma2024bisection}.

In this paper, we address these questions by focusing on one of the most widely used resampling techniques: the bootstrap\footnote{Here “bootstrap’’ refers to the statistical resampling method, distinct from the use of ``bootstrap'' in quantum many‑body physics; see \citet{poland2019conformal}.} \citep{efron1979bootstrap}. Let \(n\) denote the sample size and \(\bsX=(X_1,\dots,X_n)\) denote the observed data. For a statistic \(\hat\theta_n=f(\bsX)\), the nonparametric bootstrap estimates the sampling distribution by resampling from the empirical distribution \(\hat{\PP}_n=n^{-1}\sum_{i=1}^n\delta_{X_i}\), where $\delta_{X_i}$ is a point mass at \(X_i\). Writing
\[
H_{\mathrm{BOOT}}(z)\;=\;\Pr_*\!\big\{f(\bsX^*)\le z \mid \bsX\big\}
\]
for the \emph{ideal} (non‑Monte‑Carlo) bootstrap cumulative distribution function (CDF) (see Section~\ref{sec:classical-bootstrap} for details), the classical practice is to approximate \(H_{\mathrm{BOOT}}(z)\) by a Monte Carlo average over \(B\) replicates, which introduces approximation error of order \(B^{-1/2}\) and a computational burden that grows quickly with both \(n\) and the cost of evaluating \(f\) \citep{Babu1983,hall1986number,hall1989efficient,shi1990weak}. Scalable variants such as the Bag of Little Bootstraps \citep{kleiner2014scalable} mitigate, but do not eliminate, these limitations.

\paragraph{QBOOT in brief.}We propose the \emph{Quantum Bootstrap} (QBOOT), which replaces the classical Monte Carlo average with a \emph{single quantum amplitude} whose value is exactly \(H_{\mathrm{BOOT}}(z)\). Concretely, QBOOT (i) encodes all \(n^n\) bootstrap resamples in a uniform quantum superposition, (ii) evaluates the indicator \(g(z,\bsX^*)=\mathbf{1}\{f(\bsX^*)\le z\}\) \emph{coherently} on every resample, and (iii) uses quantum amplitude estimation (QAE) to read out the amplitude of the “accept’’ subspace, which equals \(H_{\mathrm{BOOT}}(z)\). Because QAE achieves the optimal \( \mathcal{O}(\epsilon^{-1}) \) dependence on the target accuracy \(\epsilon\) (as opposed to the classical \( \mathcal{O}(\epsilon^{-2}) \)), this yields a near‑quadratic improvement in the precision–cost trade‑off. Moreover, the construction applies to \emph{any} statistic \(f\) for which a reversible circuit implementing \(g(z,\cdot)\) can be built, turning QBOOT into a general inference primitive rather than a task‑specific speedup.


\paragraph{Contributions.}
This paper develops an end‑to‑end framework for \emph{quantum‑}\emph{powered} resampling‑based inference and makes the following specific contributions:
(i) \emph{Amplitude as ideal bootstrap value:} We show that the amplitude of a marked subspace in the QBOOT state equals the \emph{ideal} bootstrap value \(H_{\mathrm{BOOT}}(z)\), thereby removing Monte Carlo approximation from the aggregation step.
(ii) \emph{Optimal accuracy–cost scaling:} By invoking QAE on this amplitude, QBOOT attains the optimal \( \mathcal{O}(\epsilon^{-1}) \) query complexity in the desired tolerance \(\epsilon\), in contrast to the classical \( \mathcal{O}(\epsilon^{-2}) \) dependence.
(iii) \emph{A general inference primitive:} The approach applies to \emph{any} statistic whose indicator \(g(z,\bsX^*)\) admits a reversible circuit \( \Ub_g(z) \), covering common choices used for confidence intervals, hypothesis tests, and model assessment.
Beyond these core ideas, we (a) provide a statistical error analysis that separates “quantum’’ and “bootstrap’’ errors and shows how to balance them by choosing the number of QAE precision qubits; (b) establish complexity under a transparent access model that costs state preparation and reversible evaluation of \(g\); and (c) validate the method on IBM’s simulator for the sample‑mean problem, including robustness enhancements (median‑of‑measurements) that align practice with theory.


\section{Results}
\subsection{Preliminary}
\label{sec:classical-bootstrap}

Let $n\in\mathbb{N}$ denote the sample size, let $(\Omega,\mathcal{F},\PP)$ be a probability space, and let $\bsX=(X_1,\ldots,X_n)\sim\PP^{\otimes n}$ be i.i.d.\ observations. A (real-valued) statistic is a measurable map $f:\Omega^n\to\mathbb{R}$; we write
\[
\hat\theta_n \;=\; f(\bsX)
\]
for the estimator of an underlying parameter $\theta\in\Theta\subseteq\mathbb{R}$. For a threshold $z\in\mathbb{R}$, define the indicator
\[
g(z,\bx) \;=\; \mathbf{1}\!\{\, f(\bx)\le z \,\},\qquad \bx\in\Omega^n.
\]
The sampling distribution function (cdf) of $\hat\theta_n$ under $\PP$ is then
\begin{equation}
H_\PP(z) \;=\; \PP^{\otimes n}\!\big( f(\bsX)\le z \big)
\;=\; E_{\PP^{\otimes n}}\!\big[\, g(z,\bsX) \,\big].
\label{eq:H_PP_repeat}
\end{equation}

\subsubsection{Ideal nonparametric bootstrap.}
The empirical distribution of $\bsX$ is $\hat{\PP}_n = n^{-1}\sum_{i=1}^n \delta_{X_i}$, where $\delta_{X}$ denotes a point mass at $X$. A \emph{bootstrap resample} $\bsX^*=(X_1^*,\ldots,X_n^*)$ is drawn i.i.d.\ from $\hat{\PP}_n$, written $\bsX^*\sim\hat{\PP}_n^{\otimes n}$. We use $\Pr_*(\cdot\mid\bsX)$ and $E_*(\cdot\mid\bsX)$ for probability and expectation with respect to the resampling mechanism, conditional on the observed data $\bsX$. The \emph{ideal bootstrap value} is
\begin{equation}
H_{\mathrm{BOOT}}(z)
\;=\;
\Pr_*\!\big( f(\bsX^*)\le z \,\big|\, \bsX \big)
\;=\;
E_{\hat{\PP}_n^{\otimes n}}\!\big[\, g(z,\bsX^*) \,\big|\, \bsX \big].
\label{eq:H_boot_exact}
\end{equation}
If we enumerate resamples by index vectors $\ib=(i_1,\ldots,i_n)\in[n]^n$ (each $i_j\in\{1,\ldots,n\}$), then each index-resample has probability $n^{-n}$ being drawn and
\begin{equation}
H_{\mathrm{BOOT}}(z)
\;=\;
\frac{1}{n^n}\sum_{\ib\in[n]^n}
g\!\big(z,(X_{i_1},\ldots,X_{i_n})\big).
\label{eq:H_boot_np}
\end{equation}
Eq.~\eqref{eq:H_boot_np} is exact and does not require distinct observed values; duplicates are naturally accounted for by their multiplicities in $[n]^n$. Evaluating \eqref{eq:H_boot_np} is intractable for even moderate $n$ because the sum has $n^n$ terms.

\subsubsection{Monte Carlo (classical) bootstrap approximation.}
In practice, $H_{\mathrm{BOOT}}(z)$ is approximated by independently drawing $B$ bootstrap resamples $\tilde{\bsX}_1^*,\ldots,\tilde{\bsX}_B^* \sim \hat{\PP}_n^{\otimes n}$ and computing
\begin{equation}
H_{\mathrm{BOOT}}^{(B)}(z)
\;=\;
\frac{1}{B}\sum_{b=1}^B g\big(z,\tilde{\bsX}_b^*\big).
\label{eq:H_boot_MC}
\end{equation}
Because $g\in\{0,1\}$, the Monte Carlo estimator is unbiased:
\[
E_*\!\big[\,H_{\mathrm{BOOT}}^{(B)}(z)\,\big|\,\bsX\big] \;=\; H_{\mathrm{BOOT}}(z),
\]
and has conditional variance
\[
\Var_*\!\big( H_{\mathrm{BOOT}}^{(B)}(z)\,\big|\,\bsX\big)
\;=\; \frac{H_{\mathrm{BOOT}}(z)\{1-H_{\mathrm{BOOT}}(z)\}}{B}
\;\le\; \frac{1}{4B}.
\]
Thus the root-mean-square error is of order $B^{-1/2}$, and achieving accuracy $\epsilon$ (in RMSE) requires $B=\Theta(\epsilon^{-2})$ Monte Carlo replicates. We refer to this Monte Carlo approximation as the classical bootstrap (CBOOT) to contrast it with the quantum bootstrap presented in this work.

\subsubsection{Computational complexity of the classical bootstrap.}
Let $C_g(n)$ denote the classical time required to evaluate $g(z,\bx)=\mathbf{1}\{f(\bx)\le z\}$ once on a size-$n$ sample (this includes the cost of computing $f$). Generating one bootstrap resample by sampling indices with replacement costs $\Theta(n)$ random draws; hence the total cost of \eqref{eq:H_boot_MC} is
\[
\Theta\!\big( B\,C_g(n) \,+\, B\,n \big),
\]
where the dominant term is typically $B\,C_g(n)$. These baseline facts will be used in Section~\ref{sec:theory} to compare classical and quantum complexities for a fixed target accuracy.

\subsection{Quantum Bootstrap (QBOOT)}
\label{sec:qboot}

QBOOT computes the ideal bootstrap value $H_{\mathrm{BOOT}}(z)$ \emph{without} Monte Carlo approximation by (i) encoding the empirical resampling measure $\hat{\PP}_n^{\otimes n}$ as a quantum superposition, (ii) evaluating the indicator $g(z,\bsX^*)=\mathbf{1}\{f(\bsX^*)\le z\}$ coherently on every resample, and (iii) recovering the resulting amplitude by quantum amplitude estimation (QAE). Throughout, $n$ is the sample size, $\hat{\PP}_n=n^{-1}\sum_{i=1}^n\delta_{X_i}$ is the empirical distribution from Section~\ref{sec:classical-bootstrap}, and $g$ is the indicator associated with the statistic $f(\bsX)$ in \eqref{eq:H_PP_repeat}.

\subsubsection{State encoding.}
To represent the empirical resampling distribution, we index observations by $\{1,\ldots,n\}$ and use \emph{index encoding} on $m=\lceil\log_2 n\rceil$ qubits: observation $X_i$ is associated with the orthonormal basis state $\ket{i-1}$\footnote{If direct access to the numerical values $\{X_i\}$ is needed inside the circuit that computes $g$, these values are loaded into a small workspace register on demand; the index encoding keeps the resampling step simple and exact.}. A single application of a preparation unitary $\Ub_{\hat{\PP}_n}$ creates the uniform superposition
\[
\ket{\hat{\PP}_n} \;=\; \frac{1}{\sqrt{n}}\sum_{i=1}^n \ket{i-1}.
\]
Placing $n$ such registers in parallel yields the exact quantum representation of the bootstrap product measure,
\begin{equation}
\ket{\hat{\PP}_n^{\otimes n}}
\;=\; \bigotimes_{j=1}^n \ket{\hat{\PP}_n}
\;=\; \frac{1}{n^{n/2}} \sum_{\ib\in[n]^n} \ket{\ib},
\label{eq:qboot_state}
\end{equation}
where $\ib=(i_1,\ldots,i_n)$ and $\ket{\ib}=\ket{i_1-1}\otimes\cdots\otimes\ket{i_n-1}$. Measuring \eqref{eq:qboot_state} in the computational basis produces each index‑resample $\ib\in[n]^n$ with probability $n^{-n}$, matching the classical bootstrap exactly.

\subsubsection{Coherent evaluation of the indicator.}
Given $z\in\RR$, QBOOT applies a unitary
\[
\Ub_g(z):\quad \ket{\ib}\ket{0}\ \longmapsto\ \ket{\ib}\,\ket{g\!\big(z,\bsX^*(\ib)\big)},
\]
which computes $g\!\big(z,\bsX^*(\ib)\big)$ on the index-specified resample $\bsX^*(\ib)=(X_{i_1},\ldots,X_{i_n})$ and writes the result to a single \emph{label qubit}. The label qubit, together with the resample (seed) register, decomposes the post-evaluation state into orthogonal ``mean'' and ``contrast'' subspaces. Applying $\Ub_g(z)$ to \eqref{eq:qboot_state} yields a two-component superposition
\begin{equation}
\ket{\phi(z)} \;=\; \sqrt{H_{\mathrm{BOOT}}(z)}\,\ket{\phi_1} \;+\; \sqrt{1-H_{\mathrm{BOOT}}(z)}\,\ket{\phi_0},
\label{eq:mean-contrast}
\end{equation}
where $\ket{\phi_1}$ (resp.\ $\ket{\phi_0}$) is supported on label-qubit state $\ket{1}$ (resp.\ $\ket{0}$) and $\braket{\phi_1|\phi_0}=0$. Existence of such a unitary (and a constructive block-diagonal realization) follows from standard arguments and is detailed in supplementary material~\ref{app:qboot-formal}. In practice, $\Ub_g(z)$ mirrors the classical routine that evaluates $f(\cdot)$ and compares it with $z$, implemented reversibly so the workspace can be uncomputed; see supplementary material Fig.~\ref{fig:QC_mean_exmp} for the realization of $\Ub_g(z)$ for the sample-mean case.


\subsubsection{Amplitude estimation and read‑out.}
With $\Ab=\Ub_g(z)\big(\Ub_{\hat{\PP}_n^{\otimes n}}\otimes I\big)$ denoting the joint preparation of \eqref{eq:mean-contrast} from the all‑zero state, QAE constructs the Grover iterate
\[
\Qb \;=\; \Ab\, S_0\, \Ab^\dagger\, S_{\phi_1},
\]
where $S_0=I-2\ket{\mathbf{0}}\bra{\mathbf{0}}$ flips the phase of the all‑zero computational state and $S_{\phi_1}$ flips the phase on the ``mean'' subspace spanned by $\ket{\phi_1}$. The operator $\Qb$ has eigenvectors $(\ket{\phi_1}\pm i\ket{\phi_0})/\sqrt{2}$ with eigenvalues $e^{\pm i\,2\tau_z}$, where $\sin^2\tau_z=H_{\mathrm{BOOT}}(z)$. Applying Quantum Phase Estimation with $T\in\mathbb{N}$ \emph{precision qubits} produces an integer outcome $Y\in\{0,\ldots,2^T-1\}$, from which the QBOOT estimator is
\begin{equation}
H_{\mathrm{QBOOT}}(z)
\;=\;
\sin^2\!\left(\frac{\pi Y}{2^T}\right).
\label{eq:H_QBOOT_def}
\end{equation}
Standard QAE accuracy guarantees imply that the stochastic error $|H_{\mathrm{QBOOT}}(z)-H_{\mathrm{BOOT}}(z)|$ is $\mathcal{O}_p(2^{-T})$ uniformly in $z$ (see Theorem~\ref{thm:qerror}).

\subsubsection{Multiple bootstrap CDF values via parallelism.}
The above procedure estimates $H_{\mathrm{BOOT}}(z)$ for a single $z$. Suppose we aim to estimate $H_{\mathrm{BOOT}}(\cdot)$ for multiple values $\{z_d\}_{d=1}^D$ with $z_1<\cdots<z_D$. Rather than repeating the routine $D$ times, we leverage quantum parallelism by defining an operation
\[
\Ub_{g}(z_1,\ldots,z_D):\quad
\ket{\ib}\ket{0}\ \longmapsto\ \ket{\ib}\,\ket{\,D-\textstyle\sum_{d=1}^D g(z_d,\bsX^*(\ib))\,}.
\]
Consequently,
\begin{align*}
\Ub_{g}(z_1,\ldots,z_D)\big(\ket{\hat{\PP}_n^{\otimes n}}\ket{0}\big)
&= \frac{1}{n^{n/2}}\sum_{\ib\in[n]^n} \ket{\ib}\,\ket{\,D-\textstyle\sum_{d=1}^D g(z_d,\bsX^*(\ib))\,} \\
&= \sqrt{H_{\mathrm{BOOT}}(z_{d'})}\,\ket{\phi^{(d')}_1}
\;+\; \sqrt{1-H_{\mathrm{BOOT}}(z_{d'})}\,\ket{\phi^{(d')}_0},
\end{align*}
where, for each threshold $d'\in\{1,\ldots,D\}$, the two components are the normalized projections onto the events
\begin{align*}
\ket{\phi^{(d')}_1}
= &\big[H_{\mathrm{BOOT}}(z_{d'})\,n^n\big]^{-1/2}
\sum_{\ib\in[n]^n} \mathbf{1}\!\left\{D-\sum_{d=1}^D g(z_d,\bsX^*(\ib))<d'\right\}
\ket{\ib}\\
& \,\ket{\,D-\textstyle\sum_{d=1}^D g(z_d,\bsX^*(\ib))\,},\\
\ket{\phi^{(d')}_0}
= &\big[(1-H_{\mathrm{BOOT}}(z_{d'}))\,n^n\big]^{-1/2}
\sum_{\ib\in[n]^n} \mathbf{1}\!\left\{D-\sum_{d=1}^D g(z_d,\bsX^*(\ib))\ge d'\right\}
\ket{\ib}\\
& \,\ket{\,D-\textstyle\sum_{d=1}^D g(z_d,\bsX^*(\ib))\,}.
\end{align*}
Hence, we can simultaneously estimate $H_{\mathrm{BOOT}}(z_d)$ for all $d$ using a multi-dimensional amplitude estimation (MDAE) routine \footnote{The MDAE algorithm employs \(D\) measurement registers ; each of the $T$ measurement units estimates differences such as \(H_{\mathrm{BOOT}}(z_{d+1})-H_{\mathrm{BOOT}}(z_d)\); see \citep{van2021quantum}.}, while the complexity of implementing $\Ub_{g}(z_1,\ldots,z_D)$ scales only logarithmically with $D$ when Quantum Random Access Memory (QRAM) enables multi-controlled operations.



\subsection{Theoretical Results}
\label{sec:theory}

Let $H_\PP$ denote the target sampling distribution in \eqref{eq:H_PP_repeat}, let $H_{\mathrm{BOOT}}$ be the ideal nonparametric bootstrap distribution in \eqref{eq:H_boot_exact}, and let $H_{\mathrm{QBOOT}}$ be the QBOOT estimator defined in \eqref{eq:H_QBOOT_def}. For c.d.f.s $F$ and $G$, write
$
K(F,G)=\sup_{z\in\RR}|F(z)-G(z)|
$
for the Kolmogorov distance. The total error decomposes as
\begin{equation}
K\!\big(H_{\mathrm{QBOOT}}, H_\PP\big)
\;\le\;
\underbrace{K\!\big(H_{\mathrm{QBOOT}}, H_{\mathrm{BOOT}}\big)}_{\text{quantum error } \Delta_Q}
\;+\;
\underbrace{K\!\big(H_{\mathrm{BOOT}}, H_\PP\big)}_{\text{bootstrap error } \Delta_B},
\label{eq:error-decomp}
\end{equation}
where $\Delta_B$ is the bootstrap error and $\Delta_Q$ is the quantum (amplitude‑estimation) error. The stochastic orders below are taken jointly over the data $\bsX\sim\PP^{\otimes n}$ and, when relevant, over the internal measurement randomness of QAE.
We first characterize the quantum error $\Delta_Q$ and then review the bootstrap error $\Delta_B$. The balance between these two terms is a critical guide for quantum circuit design and resource allocation: choosing $T$ too small makes $\Delta_Q$ dominate, while choosing $T$ too large wastes quantum resources beyond what the bootstrap accuracy requires.

\subsubsection{Quantum error.}
\begin{theorem}[Exact distribution of QBOOT]
\label{thm:exact_dist}
Fix $T\in\mathbb{N}$ and $z\in\mathbb{R}$. 
Conditional on $H_{\mathrm{BOOT}}(z)$ (equivalently, on the observed data), the QBOOT estimator admits the exact representation
\begin{equation}
H_{\mathrm{QBOOT}}(z) \;\overset{d}{=}\; \sin^2\!\left(\frac{\pi Y_z}{2^T}\right),
\label{eq:exact_representation}
\end{equation}
where $Y_z\in\{0,1,\dots,2^T-1\}$ has probability mass function
\begin{equation}
\Pr(Y_z = l)
=
\frac{\sin^2(2^T\tau_z)}{2^{2T+1}}
\left[
\csc^2\!\left(\tau_z - \frac{l\pi}{2^T}\right)
+
\csc^2\!\left(\tau_z + \frac{l\pi}{2^T}\right)
\right],
\qquad l=0,1,\dots,2^T-1,
\label{eq:exact_pmf}
\end{equation}
with $\tau_z = \arcsin\sqrt{H_{\mathrm{BOOT}}(z)}$
for $\tau_z \notin \{\frac{m\pi}{2^T}: m\in\mathbb{Z}\}$, with \eqref{eq:exact_pmf} understood by continuous extension at those grid points.
In particular, if $2^T\tau_z/\pi\in\mathbb{Z}$ (equivalently $p_z=\sin^2(m\pi/2^T)$ for some integer $m$), then $Y_z$ is degenerate and
\[
\Pr\!\left(H_{\mathrm{QBOOT}}(z) = H_{\mathrm{BOOT}}(z)\right)=1.
\]
\end{theorem}

 Since $\sin^2$ is $1$-Lipschitz, we can derive the stochastic bound of the quantum error,
\begin{theorem}[Quantum error stochastic bound]
\label{thm:qerror}
Let $T$ be the number of precision qubits used by QAE. Then
\[
\Delta_Q \;=\; K\!\big(H_{\mathrm{QBOOT}}, H_{\mathrm{BOOT}}\big) \;=\; \mathcal{O}_p(2^{-T}),
\]
for $\forall z\in\RR$.
\end{theorem}


\begin{theorem}[Bias and MSE]
\label{thm:biasmse}
For any fixed $z\in\RR$ and $T\in\mathbb{N}$,
\begin{align}
\mathbb{E}\!\big[H_{\mathrm{QBOOT}}(z)\,\big|\,\bsX\big] - H_{\mathrm{BOOT}}(z)
&= 2^{-2T}\sin^2(2^T\tau_z)\sum_{l=0}^{2^T-1}\frac{\sin(2^{-T}l\pi+\tau_z)}{\sin(2^{-T}l\pi-\tau_z)}, \label{eq:bias}\\
\mathbb{E}\!\Big[\!\big(H_{\mathrm{QBOOT}}(z)-H_{\mathrm{BOOT}}(z)\big)^2 \,\Big|\,\bsX\Big]
&= 2^{-2T}\sin^2(2^T\tau_z)\sum_{l=0}^{2^T-1}\sin^2(2^{-T}l\pi+\tau_z). \label{eq:mse}
\end{align}
\end{theorem}
 
\begin{figure}[!htb]
  \centering
  \includegraphics[width=1\textwidth]{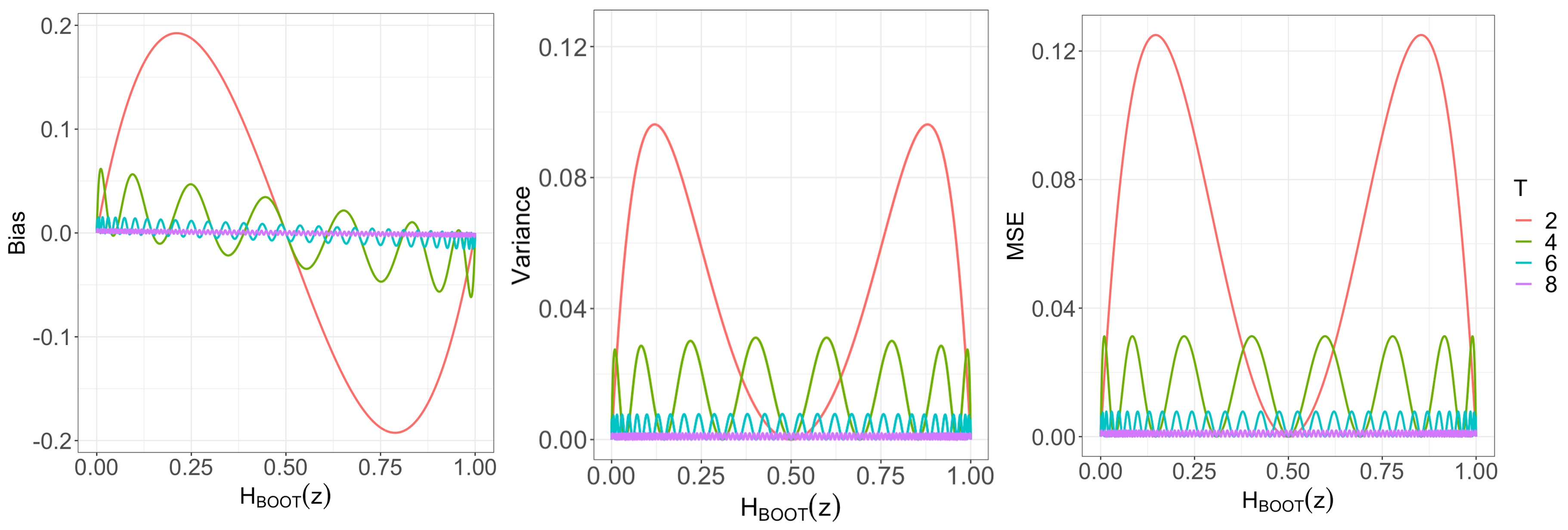}
  \caption{Bias, variance, and MSE of $H_{\mathrm{QBOOT}}(z)$ as functions of $H_{\mathrm{BOOT}}(z)$ for $T\in\{2,4,6,8\}$, showing oscillations and the vanishing of error when $2^T\tau_z/\pi\in\mathbb{Z}$.}
  \label{fig:T_selection}
\end{figure}

The bias and MSE depend on $z$ through $\tau_z$ and exhibit a mild oscillatory pattern with exact zeros when $2^T\tau_z/\pi\in\mathbb{Z}$; see Fig.~\ref{fig:T_selection}. These oscillations explain the non‑monotone but \emph{overall} $2^{-T}$ decay observed empirically. Using this theorem, we can derive uniform asymptotic bounds for the bias and MSE of the quantum error.
\begin{corollary}[Uniform asymptotics]
\label{cor:asymptotic}
As $T\to\infty$, uniformly in $z\in\RR$,
\[
\big|\mathrm{Bias}\big| \;=\; \mathcal{O}(2^{-T}),
\qquad
\mathrm{MSE} \;=\; \mathcal{O}(2^{-T}).
\]
\end{corollary}

\subsubsection{Work (gate) complexity.}
Let $Q_g(n)$ be the gate complexity of a single call to the unitary $\Ub_g(z)$ that evaluates $g(z,\bsX^*)=\mathbf{1}\{f(\bsX^*)\le z\}$ (Section~\ref{sec:qboot}); let $Q_{\mathrm{enc}}(n)$ be the cost of preparing the $n$ resampling registers (index encoding), which is $\mathcal{O}(n\log n)$ gates, as discussed in detail in Section \ref{app:implementation} of the supplementary material. One run of QAE with $T$ precision qubits uses $\Theta(2^T)$ controlled applications of the Grover iterate $\Qb$, and each $\Qb$ invokes $\Ab$ and $\Ab^\dagger$ once (hence calls $\Ub_g(z)$ and $\Ub_g(z)^\dagger$ once). Therefore, to achieve the accuracy in Theorem~\ref{thm:qerror}, the work complexity is
\begin{equation}
\text{Work}(\text{QBOOT},\epsilon\asymp 2^{-T})
\;=\;
\mathcal{O}\!\big(2^T\,\big(Q_g(n)\,\vee\,n \log n(n)\big)\big),
\label{eq:qboot-work}
\end{equation}
where $a\vee b=\max\{a,b\}$. For the classical Monte Carlo bootstrap (CBOOT), attaining error $\epsilon$ requires $B=\Theta(\epsilon^{-2})$ resamples, at cost
\begin{equation}
\text{Work}(\text{CBOOT},\epsilon)
\;=\;
\mathcal{O}\!\big(\epsilon^{-2}\,\big(C_g(n)\,\vee\,n\log n\big)\big),
\label{eq:cboot-work}
\end{equation}
where $C_g(n)$ is the classical cost of evaluating $g$ once and $n\log n$ accounts for generating each resample.

{In theory, $Q_g(n) \lesssim C_g(n)$. For many algorithms, the cost can even be much lower. See the supplementary material Table \ref{tab:Q_g_vs_C_g} for a detailed comparison of different tasks.}

The following theorem presents the gate complexity of QBOOT.
\begin{theorem}[QBOOT gate complexity]
\label{thm:complexity}
To achieve $\Delta_Q=\mathcal{O}_p(2^{-T})$, QBOOT requires $\mathcal{O}\!\big(2^T\,(Q_g(n)\vee n\log n)\big)$ gates. In terms of a target tolerance $\epsilon\asymp 2^{-T}$, this equals $\mathcal{O}(\epsilon^{-1}\,(Q_g\vee n\log n))$, whereas CBOOT requires $\mathcal{O}(\epsilon^{-2}\,(C_g\vee n\log n))$ operations.
\end{theorem}

{%
\subsubsection{Qubit complexity of QBOOT.}


Having established the queries to $\Ub_g(z)$ and gate complexity, we now explicitly analyze the total qubit count.

The QBOOT circuit has four distinct qubit components: (1) The \emph{bootstrap register} uses $n\lceil\log_2 n\rceil$ qubits to
hold $n$ index registers of $\lceil\log_2 n\rceil$ qubits each.
(2) The \emph{statistic register} stores the computed value of $f(\bsX^*)$.
When $f$ is a discrete statistic taking at most $V$ distinct values
(e.g., the sample mean of $n$ draws from $\{0,\ldots,n-1\}$ has
$V=n(n-1)+1$ possible values), this requires
$\lceil\log_2(V+1)\rceil$ qubits.
For a continuous statistic standardized to $[0,1]$, one could discretize to
allow a rounding error of at most $1/V$;
choosing $V \gg n$ keeps this rounding error negligible compared to the bootstrap error $\Delta_B$.
(3) The \emph{label qubit} is a single qubit that stores the binary indicator $g(z,\bsX^*)\in\{0,1\}$ after the unitary $\Ub_g(z)$ is applied.
(4) The \emph{QAE register} contains $T$ precision qubits used by the quantum phase estimation step of QAE to estimate
$\sqrt{H_{\mathrm{BOOT}}(z)}$ to accuracy $\mathcal{O}(2^{-T})$.

Combining all components, the QBOOT circuit with $T$ precision qubits
requires
\begin{equation}
  N_{\mathrm{qubit}}
  \;=\;
  \underbrace{n\lceil\log_2 n\rceil}_{\text{bootstrap register}}
  +\,
  \underbrace{\lceil\log_2(V+1)\rceil}_{\text{statistic register}}
  +\,
  \underbrace{1}_{\text{label qubit}}
  +\,
  \underbrace{T}_{\text{QAE register}}
  +\,W_f,
  \label{eq:qubit-count}
\end{equation}
where $W_f$ is the count of ancillary qubits  for holding intermediate values during reversible
computation. $W_f$ depends on the algorithm for computing the statistic $f$ and is typically considered to be bounded by a constant, i.e., does not depend on $n$.

The dominant term in the qubit complexity, $n\lceil\log_2 n \rceil$, is generically the bootstrap register,
which must hold all $n$ index registers simultaneously to maintain
coherence across all $n^n$ resamples.
This limits near-term applicability to modest sample sizes.
A natural mitigation is the \emph{$m$-out-of-$n$ bootstrap}
\citep{politis1994large}, replacing the $n$-register system with $m\ll n$
registers, reducing the qubit count to $\mathcal{O}(m\log n)$ at the
cost of a slower bootstrap convergence rate
$\Delta_B=\mathcal{O}_p(m^{-1/2})$.
}

\subsubsection{Balancing stochastic errors and quantum resource allocation.}
Under standard regularity for the \emph{smoothness} of $f$ (e.g., considering Hadamard differentiable statistics such as the sample mean, variance, and smooth $M$-estimators; see \citealp{bickel1981some,efron1994introduction,hall2013bootstrap}), the bootstrap is second‑order accurate:
\begin{equation}
\Delta_B \;=\; \mathcal{O}_p(n^{-1}).
\label{eq:bootstrap-second-order}
\end{equation}
This bound is uniform in $z$ on continuity points of $H_\PP$.
Combining \eqref{eq:bootstrap-second-order} and Theorem~\ref{thm:qerror}, the total Kolmogorov error in \eqref{eq:error-decomp} satisfies
$
K(H_{\mathrm{QBOOT}},H_\PP)
=\mathcal{O}_p(2^{-T})+\mathcal{O}_p(n^{-1}).
$
Choosing $T$ to balance these terms, e.g.\ $2^{-T}\asymp n^{-1}$ (so $T\asymp \log_2 n$), yields the overall rate $\mathcal{O}_p(n^{-1})$ with quantum work
\[
\text{Work}(\text{QBOOT}) \;=\; \mathcal{O}\!\big(n\,(Q_g(n)\vee n \log n(n))\big),
\]
whereas CBOOT requires $B\asymp n^2$ replicates and hence $\mathcal{O}(n^2\,(C_g(n)\vee n\log n))$ work. When $Q_g(n)\lesssim C_g(n)$, this delivers at least an $\Omega(n)$ asymptotic speedup at matched statistical accuracy.

For \emph{non-smooth statistics}, e.g.\ sample quantiles or
spectral statistics,
$\Delta_B=\mathcal{O}_p(n^{-1/2})$
\citep{bickel1981some,efron1994introduction}.
In that case, balancing with $\Delta_Q$ sets
$T\asymp\frac{1}{2}\log_2 n$ (so $2^{-T}\asymp n^{-1/2}$),
giving work $\mathcal{O}(\sqrt{n}\,(Q_g\vee n \log n))$
for QBOOT versus $\mathcal{O}(n\,(C_g\vee n\log n))$ for CBOOT
— a speedup of $\Omega(\sqrt{n})$ even in the non-smooth regime.

In Table \ref{tab:complexity}, we compare the work–error trade‑offs for target tolerance $\epsilon\asymp 2^{-T}$ between QBOOT and CBOOT in detail.

\begin{table*}
\caption{Work–error trade‑offs for target tolerance $\epsilon\asymp 2^{-T}$ (one value of $z$).}
\begin{ruledtabular}
\begin{tabular}{lcc}
 & QBOOT & CBOOT \\
\midrule
Error (per $z$) vs.\ work & $\epsilon\asymp \text{work}^{-1}$ & $\epsilon\asymp \text{work}^{-1/2}$ \\
\hline
Sampling over resamples & $n^n$ (in superposition) & $B=\Theta(\epsilon^{-2})$ Monte Carlo draws \\
\hline
Time (per $z$) & $\mathcal{O}(\epsilon^{-1}(Q_g\vee n \log n))$ & $\mathcal{O}(\epsilon^{-2}(C_g\vee n\log n))$ \\
\hline
Space (qubits / works) & $\Theta(n\log n)+T$ qubits & $\Theta(Bn)$ works \\
\end{tabular}
\label{tab:complexity}
\end{ruledtabular}
\end{table*}



\subsection{Numerical Experiments}
\label{sec:num_exp}

We evaluate QBOOT on the IBM Quantum Experience simulator via Qiskit.\footnote{\url{https://qiskit.org/}}
Unless stated otherwise, all runs use a noiseless backend.

\subsubsection{Data and statistic.}
We fix the observed sample
\[
\bsX=\{0,1,2,3\},\qquad n=4,
\]
and consider the (with-replacement) bootstrap of the sample mean
\[
\bar X^{*} \;=\; \frac{1}{n}\sum_{j=1}^{n} X^{*}_j,\qquad n=4.
\]
Our target is the bootstrap CDF at \(z=1.25\):
\[
H_{\mathrm{BOOT}}(z)\;=\;\Pr\!\big\{\bar X^{*}\le z\,\big|\,\bsX\big\}\quad\text{evaluated at }z=1.25.
\]

\subsubsection{Ground truth by enumeration.}
For \(n=4\) and alphabet \(\{0,1,2,3\}\), there are \(n^n=4^4=256\) equally likely resamples
\((x_1,\ldots,x_4)\in\{0,1,2,3\}^4\).
Since \(\bar X^{*}\le z\) is equivalent to \(\sum_{j=1}^{4} x_j \le n z = 5\), the ideal bootstrap value is
\begin{eqnarray*}
    H_{\mathrm{BOOT}}(1.25)
\;&=&\;\frac{\#\{(x_1,\ldots,x_4)\in\{0,1,2,3\}^4:\; x_1+\cdots+x_4 \le 5\}}{256}
\;\\
&=&\;\frac{106}{256}\;\approx\;0.4140625.
\end{eqnarray*}

We use this enumerated value as the ground truth when reporting QBOOT errors.

\subsubsection{Circuit layout (QBOOT).}
The \emph{bootstrap sample register} contains $n=4$ independent subsystems, each using $m=\lceil \log_2 n \rceil=2$ qubits to index $\{0,1,2,3\}$; parallel Hadamard gates prepare the uniform superposition for each subsystem. 
The \emph{statistic register} uses $4$ qubits to accumulate the sum $\sum_{i=1}^4 X_i^* \in \{0,\ldots,12\}$, and a single \emph{label qubit} stores the indicator $g(z,\bsX^*)=\mathbf{1}\{\bar{X}^*\le z\}$. 
Thus, the encoding and processing units use $8+4+1=13$ qubits. 
The \emph{QAE measurement register} uses $T\in\{4,5,\ldots,10\}$ precision qubits.
In total, the circuit uses $13+T$ qubits. 
The comparator implements $\sum_{i=1}^4 X_i^* \le 5$ (equivalently $\bar{X}^* \le 1.25$) and flips the label qubit accordingly; see supplementary material Fig.~\ref{fig:QC_mean_exmp} for the gate-level realization.

\subsubsection{Cost matching and replication protocol.}
Let $Q_g$ denote the gate complexity of the statistic-evaluation unitary $\Ub_g(z)$, and let $C_g$ denote the classical cost of evaluating $g(z,\cdot)$ on one bootstrap resample. 
For the sample mean, $Q_g \asymp C_g \asymp n$.
Under this scaling, a single QAE run with $T$ precision qubits makes $\mathcal{O}(2^T)$ calls to $\Ub_g(z)$, giving an overall cost $\mathrm{Cost}(\mathrm{QBOOT})=\mathcal{O}(2^T\,n\log n)$. 
To compare fairly, the classical bootstrap (CBOOT) uses $B=2^T$ resamples so that $\mathrm{Cost}(\mathrm{CBOOT})=\mathcal{O}(B\,n\log n)$ matches the QBOOT cost up to constants.

For each $T\in\{4,\ldots,10\}$ we run $5{,}000$ independent replications. 
In each replication, QBOOT executes one canonical-QAE run and returns the estimate
\[
H_{\mathrm{QBOOT}}(z)=\sin^2\!\left(\frac{Y\pi}{2^T}\right), 
\]
where $Y\in\{0,\ldots,2^T-1\}$ is the QAE outcome.
CBOOT draws $B=2^T$ i.i.d.\ resamples from $\hat{\PP}_n$ and returns
$
H_{\mathrm{BOOT}}^{(B)}(z)=B^{-1}\sum_{b=1}^B \mathbf{1}\{\bar{X}^{*(b)}\le z\}.
$

\subsubsection{Error metrics.}
For a generic estimator $\widehat H(z)$ (QBOOT or CBOOT), we compute the absolute error
$
\mathrm{Err} = \big|\widehat H(z) - H_{\mathrm{BOOT}}(z)\big|
$
with $H_{\mathrm{BOOT}}(z)=106/256$ as the ground truth.
We summarize the distribution of $\log(\mathrm{Err})$ across replications and also report three scalar summaries as functions of $T$: mean absolute error (MAE), $\sqrt{\mathrm{MSE}}$, and median absolute error.

\subsubsection{Main findings (single QAE measurement).}
\begin{figure}[!htb]
  \centering
  \includegraphics[width=1\textwidth]{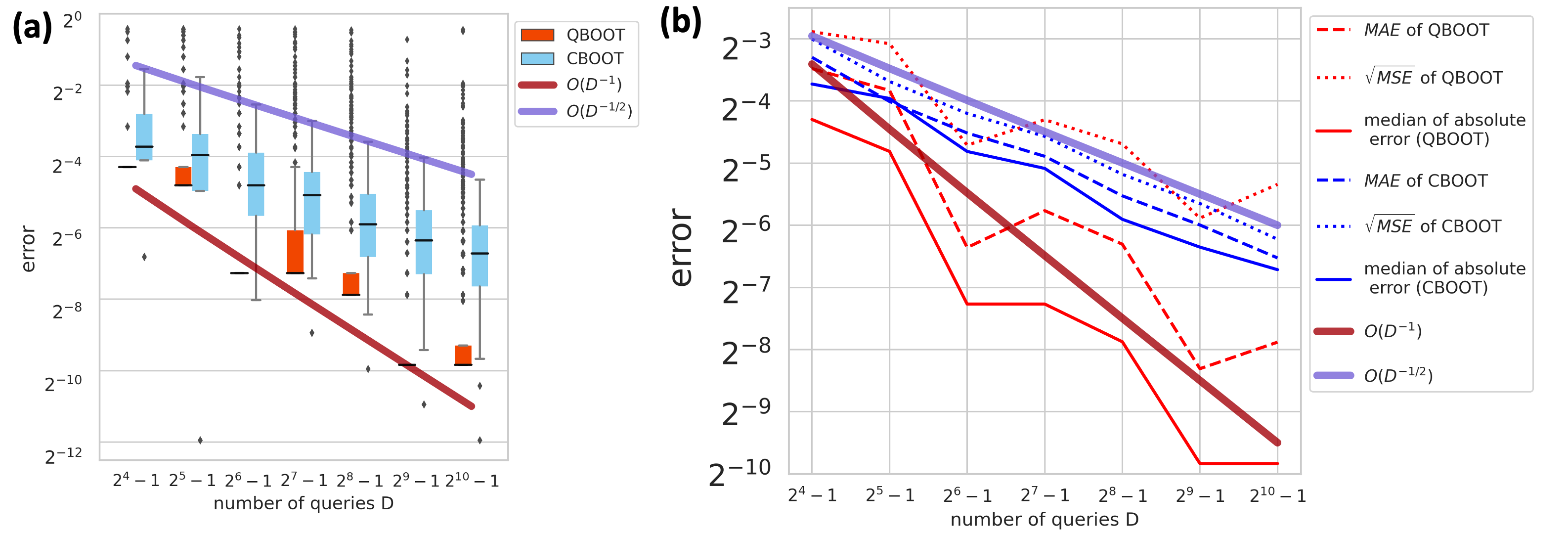}
  \caption{QBOOT vs.\ CBOOT at matched cost. 
  (a) Boxplots of $\log$ absolute errors vs.\ $T$, with theoretical guides $O(2^{-T})$ (QBOOT, red) and $O(2^{-T/2})$ (CBOOT, blue).
  (b) MAE, $\sqrt{\mathrm{MSE}}$, and median absolute error vs.\ $T$.}
  \label{fig:qc_simu}
\end{figure}

Figure~\ref{fig:qc_simu}(a) displays the log absolute error distribution across $T$. 
The QBOOT error decreases approximately at the theoretically predicted $O(2^{-T})$ rate (Theorem~\ref{thm:qerror}), while CBOOT follows the $O(2^{-T/2})$ rate implied by Monte Carlo fluctuations. 
For instance, at $T=6$, CBOOT requires on the order of $B\gtrsim 2^{10}$ resamples to match the accuracy of QBOOT with $T=6$ (which uses only $2^6$ coherent queries).
Figure~\ref{fig:qc_simu}(b) shows that QBOOT achieves uniformly smaller MAE and median absolute error than CBOOT at matched cost, whereas the $\sqrt{\mathrm{MSE}}$ curves are closer due to occasional large QAE errors (see below).

\subsubsection{Outliers and their mitigation.}
\begin{figure}[!htb]
  \centering
  \includegraphics[width=1\textwidth]{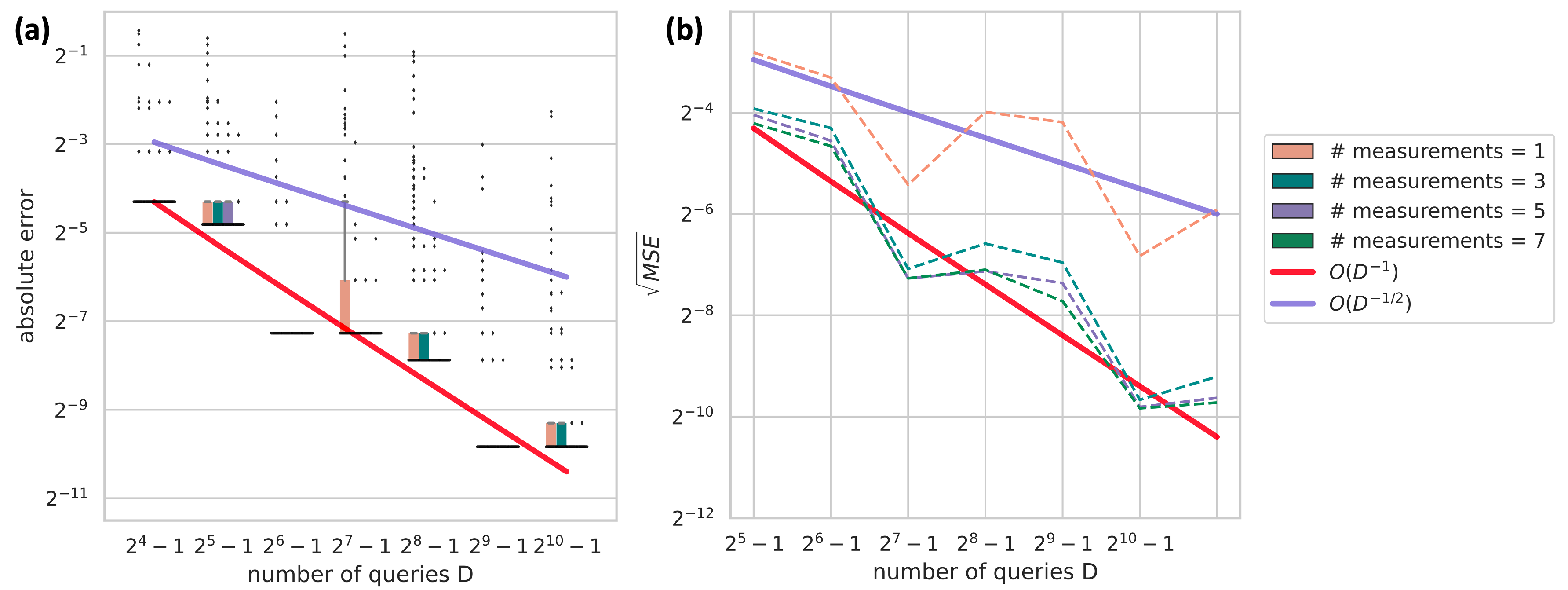}
  \caption{Robust aggregation for QBOOT. 
  (a) Absolute error boxplots vs.\ $T$ for $M\in\{1,3,5,7\}$ repeated QAE measurements; the median across $M$ runs is reported per replication. 
  (b) $\sqrt{\mathrm{MSE}}$ vs.\ $T$; robust aggregation (e.g., $M=5$) closely tracks the $O(2^{-T})$ guide.}
  \label{fig:qc_simu_median}
\end{figure}

The canonical QAE estimator exhibits a small fraction of large errors caused by phase \emph{aliasing} (the discretization of the phase grid at resolution $2^{-T}$) and the oscillatory dependence of the error on $\sin^2\tau_z$ (Section~\ref{sec:theory}). 
A standard remedy is to repeat the QAE call and aggregate robustly.
We therefore repeat QAE $M\in\{1,3,5,7\}$ times per replication and report the \emph{median-of-$M$} QBOOT estimates. 
Figure~\ref{fig:qc_simu_median}(a) shows that increasing $M$ sharply reduces both the frequency and the magnitude of outliers. 
Figure~\ref{fig:qc_simu_median}(b) shows that $\sqrt{\mathrm{MSE}}$ approaches the ideal $O(2^{-T})$ slope with as few as $M\approx 5$ repeats, while preserving QBOOT’s advantage in MAE and median error.

On this controlled example, QBOOT attains the expected $O(2^{-T})$ accuracy in estimating the \emph{ideal} bootstrap value, whereas CBOOT, run at matched computational cost, exhibits the slower $O(2^{-T/2})$ decay. 
The occasional large errors of the canonical QAE estimator are well documented theoretically and are effectively mitigated by a small number of repeated measurements with median aggregation, which brings the empirical $\sqrt{\mathrm{MSE}}$ close to the $O(2^{-T})$ benchmark.

{
\subsection{Theoretical analysis on Median-of-$M$ QBOOT}
As noted in Section~\ref{sec:num_exp}, canonical QAE
occasionally produces large errors.
Our remedy is to run QAE independently $M$ times and
to take the median,
\begin{equation}
  H_{\mathrm{QBOOT}}^{(M)}(z)
  \;=\;
  \mathrm{median}\!\left\{
    H_{\mathrm{QBOOT}}^{(1)}(z),\ldots,H_{\mathrm{QBOOT}}^{(M)}(z)
  \right\},
  \label{eq:median-estimator}
\end{equation}
where each $H_{\mathrm{QBOOT}}^{(m)}(z)$ is an independent
QAE runs with $T$ precision qubits.
\begin{lemma}[Single-run error bound]
\label{prop:single-run}
Let $\varepsilon_T=\pi/2^T$ and let $l^*=\lfloor 2^T\tau_z/\pi\rfloor$. By Theorem~\ref{thm:exact_dist}, the single-run error can be bounded by 
\begin{equation}
  q_z \;:=\; \Pr\!\bigl(|H_{\mathrm{QBOOT}}(z)-H_{\mathrm{BOOT}}(z)|
  \le\varepsilon_T\bigr)
  \;\ge\; \frac{8}{\pi^2} \;>\; \frac{1}{2},
  \label{eq:exact-success}
\end{equation}
uniformly in $z\in\mathbb{R}$, with equality holding only when $2^T\tau_z/\pi$ is a half-integer (worst-case phase).
See Supplementary Section~\ref{app:proof-median} for the derivation.
\end{lemma}

We can then derive the error bound of the Median-of-M QBOOT.

\begin{theorem}[Median-of-$M$ error bound]
\label{thm:median-M}
Let $\rho = 32(\pi^2-8)/\pi^4 \approx 0.614$.
For any $M\in\mathbb{N}$ and $z\in\mathbb{R}$,
\begin{equation}
  \Pr\!\left(|H_{\mathrm{QBOOT}}^{(M)}(z)-H_{\mathrm{BOOT}}(z)|
    >\frac{\pi}{2^T}\right)
  \;\le\; \rho^{M/2}.
  \label{eq:median-bound}
\end{equation}

\end{theorem}
In particular, when
\begin{equation}
  M \;\ge\; \frac{2\log(1/\delta)}{\log(1/\rho)}
  \;\approx\; 4.11\,\log(1/\delta),
  \label{eq:M-choice}
\end{equation}
the failure probability $\le\delta$.


\begin{theorem}[Median estimate preserves the quadratic speedup]
\label{thm:cost-median}
Let $\varepsilon_T=\pi/2^T$ and $\delta\in(0,1)$ be the target error and the target failure probability.
\begin{description}
  \item[QBOOT-$M$] using $M=\lceil\log(1/\delta)/(2c_0)\rceil$
    independent QAE runs each with $T$ precision qubits,
    \begin{equation}
      \mathrm{Work}(\mathrm{QBOOT\text{-}M},\varepsilon_T,\delta)
      =\mathcal{O}\!\bigl(\varepsilon_T^{-1}\log(1/\delta)\,(Q_g\vee n\log n )\bigr).
      \label{eq:qboot-M-work}
    \end{equation}
  \item[CBOOT] Hoeffding's inequality requires
    $B\ge(2\varepsilon_T^2)^{-1}\log(2/\delta)$ resamples to
    achieve error $\varepsilon_T$ and failure probability $\delta$:
    \begin{equation}
      \mathrm{Work}(\mathrm{CBOOT},\varepsilon_T,\delta)
      =\mathcal{O}\!\bigl(\varepsilon_T^{-2}\log(1/\delta)\,(C_g\vee n\log n)\bigr).
      \label{eq:cboot-delta-work}
    \end{equation}
  Therefore, under $Q_g\lesssim C_g$, the ratio
    $\mathrm{Work(CBOOT)}/\mathrm{Work(QBOOT\text{-}M)}
    =\Omega(\varepsilon_T^{-1})=\Omega(2^T)$.
    The near-quadratic speedup is therefore fully preserved
    after median aggregation.
\end{description}
\end{theorem}

\begin{remark}
For $\delta=0.05$, Theorem~\ref{thm:cost-median} implies
$M\lesssim13$ repeats suffice — a constant-factor overhead
that does not affect the asymptotic speedup over CBOOT.
The empirical observation that $M=5$--$7$ repeats already
suffice at $T\ge6$ (Section~\ref{sec:num_exp}) is consistent
with $\rho$ being a worst-case bound: for most $\tau_z$,
the exact single-run success probability $q_z$ (Lemma~\ref{prop:single-run})
exceeds $8/\pi^2$, making $\rho^{M/2}$ a conservative
overestimate of the true failure probability of the median-of-$M$ QBOOT.
\end{remark}
}
\section{Discussion}

QBOOT offers a principled route to \emph{quantum‑powered} resampling–based inference. 
By preparing a coherent superposition over all $n^n$ bootstrap resamples and evaluating the indicator $g(z,\bsX^*)=\mathbf{1}\{\!f(\bsX^*)\le z\!\}$ in parallel, QBOOT removes Monte Carlo approximation from the bootstrap step and replaces it with a controllable quantum estimation error governed by the number $T$ of precision qubits. 
The analysis in Section~\ref{sec:theory} shows that the quantum error obeys $K(H_{\mathrm{QBOOT}},H_{\mathrm{BOOT}})=\mathcal{O}_p(2^{-T})$ (Theorem~\ref{thm:qerror}), while the total computational cost scales as $\mathcal{O}\!\big(2^{T}(Q_g\vee Q_{\mathrm{enc}})\big)$ (Theorem~\ref{thm:complexity}). 
Balancing this with the classical bootstrap error suggests the practical choice $2^{-T}\asymp n^{-1}$, which yields overall cost $\mathcal{O}(n\,Q_g)$ and at least a linear‑in‑$n$ advantage over the classical bootstrap when $Q_g\lesssim C_g$. 
In our experiments on the sample‑mean problem, QBOOT exhibits the predicted $O(2^{-T})$ decay in absolute error and dominates the classical bootstrap (CBOOT) in mean and median absolute error for matched cost; modest repetition with median aggregation suppresses the occasional large QAE errors and restores $\sqrt{\mathrm{MSE}}$ to the $O(2^{-T})$ trend. Table \ref{tab:qboot_vs_cboot} provides a detailed theoretical and practical comparison of QBOOT and CBOOT.

There are two important distinctions from classical practice. 
First, QBOOT targets the \emph{ideal} bootstrap quantity $H_{\mathrm{BOOT}}(z)$ rather than a Monte Carlo approximation $H_{\mathrm{BOOT}}^{(B)}(z)$. 
Second, the accuracy–cost trade‑off is explicit and tunable: halving the quantum error requires one additional precision qubit $T\mapsto T+1$ (i.e., a constant‑factor increase in queries), whereas the classical Monte Carlo error requires a four‑fold increase in the number of resamples $B$.

The performance guarantees rely on standard but explicit assumptions: 
(i) an orthogonal encoding of the observed values into computational basis states so that resamples are perfectly distinguishable; 
(ii) a unitary implementation of the statistic oracle $\Ub_g(z)$ with gate complexity $Q_g$ that is comparable to the classical evaluation cost $C_g$; and 
(iii) state preparation with cost $n \log n=\mathcal{O}(n\log n)$ using index encoding (Section~\ref{sec:qboot}). 
These assumptions hold for many statistics that can be computed with reversible arithmetic (e.g., sums, ranks, comparisons) and can be tightened further when quantum subroutines offer genuine savings (e.g., Grover‑type routines for order statistics or search, quantum kernels for certain similarity measures). 
As with any amplitude‑estimation method, finite‑precision effects induce an oscillatory error profile; our empirical results show that a small number of repeated QAE runs with median aggregation largely removes their impact on MSE without altering the asymptotic rate.

Canonical QAE utilizes controlled powers of a Grover iterate and an inverse QFT, which can be demanding on noisy devices. 
In near‑term settings, it is natural to replace the phase‑estimation primitive by shallower amplitude‑estimation variants and to use error‑mitigated circuits for the arithmetic and comparator blocks. 
Our study utilizes a simulator to isolate algorithmic behavior; hardware deployment will benefit from depth-reduction strategies (e.g., iterative amplitude estimation, decomposition of comparators into low-T gates) and the robust aggregation scheme we have evaluated.

QBOOT’s design is modular at the resampling, statistic, and estimation layers. 
Alternative resampling schemes—parametric, smoothed, Bayesian, $m$‑out‑of‑$n$—amount to changing the state‑preparation unitary. 
The same framework extends beyond distributional estimation to hypothesis testing (quantum permutation tests) and model assessment (quantum cross‑validation), where the label qubit encodes the relevant indicator (or loss threshold) and QAE provides the aggregate. 
When multiple thresholds $z$ are needed, one can reuse the bootstrap superposition and evaluate multiple comparators sequentially, amortizing the state-preparation cost across the grid.

In summary, QBOOT preserves the asymptotic properties of the ideal bootstrap while delivering a clear computational advantage in the accuracy–cost trade‑off. 
Although current NISQ devices limit problem sizes, rapid improvements in quantum volume and error correction should expand the feasible regime. 
We anticipate that QBOOT and related quantum resampling algorithms will become practical tools for large‑scale, model‑free statistical inference, especially for statistics whose classical evaluation is costly and for workflows that require many distributional queries $z$.

\begin{table*}[t]
\caption{Theoretical and practical comparison of QBOOT and CBOOT. Error\\ rates refer to a target tolerance $\epsilon$ for estimating $H_{\mathrm{BOOT}}(z)$; complexity is \\ assumed to be $Q_g\asymp C_g$ for comparability.}
\label{tab:qboot_vs_cboot}
\begin{ruledtabular}
\begin{tabular}{p{0.15\linewidth}p{0.38\linewidth}p{0.39\linewidth}}
 & \textbf{QBOOT} & \textbf{CBOOT} \\
\midrule
Error for $H_{\mathrm{BOOT}}(z)$ & $\mathcal{O}(\epsilon)$ via QAE (queries scale as $\epsilon^{-1}$) & $\mathcal{O}(B^{-1/2})$ with $B$ resamples (cost scales as $\epsilon^{-2}$) \\
\hline
Bootstrap sampling & All $n^n$ resamples in superposition (no MC error) & $B$ Monte Carlo resamples (approximate) \\
\hline 
Complexity (per $z$) & $\mathcal{O}(\epsilon^{-1}Q_g)$; prep. $\mathcal{O}(n\log n)$ & $\mathcal{O}(\epsilon^{-2}C_g)$;  prep. $\mathcal{O}(n\log n)$ \\
\hline
Balancing errors & $2^{-T}\!\asymp n^{-1}\Rightarrow \mathcal{O}(nQ_g)$ overall & $B\!\asymp n^2\Rightarrow \mathcal{O}(n^2C_g)$ overall \\
\hline
Empirical (this study) & $O(2^{-T})$ rate; lower MAE/median at matched cost; MSE restored by small repeats & $O(2^{-T/2})$ rate; larger MAE/median at matched cost \\
\hline
Flexibility & Directly adapts to other resampling and inference tasks & Widely used; no inherent quantum speedup \\
\end{tabular}
\end{ruledtabular}
\end{table*}

\section*{Acknowledgments}
\paragraph*{Funding:}
This work was partially supported by the U.S. National Science Foundation [DMS-1925066, DMS-1903226, DMS-2124493, DMS-2311297, DMS-2319279, DMS-2318809] and the National Institutes of Health [NIH R01GM152814].
\paragraph*{Author contributions:}
P.M., Y.C., and W.Z. designed research; P.M., Y.C., and W.Z. performed research; Y.C. implemented the algorithm, established the theory, and conducted the empirical analysis. All authors participated in writing the paper.
\paragraph*{Competing interests:}
There are no competing interests to declare.

\paragraph*{Data, Materials, and Code Availability}
The source code and experimental data supporting the QBOOT algorithm are available at \url{https://github.com/StatCYK/quantum-bootstrap}.



\bibliography{ref}

\newpage

\appendix
\setcounter{page}{1} 
\setcounter{figure}{0}
\renewcommand{\thefigure}{S.\arabic{figure}}
\renewcommand{\thesection}{S.\arabic{section}}
\renewcommand{\theequation}{S.\arabic{equation}}
\renewcommand{\thealgorithm}{S.\arabic{algorithm}}

\begin{center}
  {\Large\bf Supplementary Material for  ``Quantum Statistical Bootstrap''}  
\end{center}

\section{Notation and Quantum Preliminaries}
\label{app:quantum-prelims}

\paragraph{Dirac notation and registers.}
The computational basis for a single qubit is $\{\ket{0},\ket{1}\}$. A qubit state is $\ket{\psi}=\alpha_0\ket{0}+\alpha_1\ket{1}$ with $|\alpha_0|^2+|\alpha_1|^2=1$.
A register of $p$ qubits is the tensor product space $\cH_2^{\otimes p}$ and supports superpositions over $2^p$ basis states.
We use $\bra{\psi}$ for the conjugate transpose of $\ket{\psi}$ and $\braket{\phi|\psi}$ for the inner product.

\paragraph{Unitary evolution and control.}
A linear map $U$ on a finite‑dimensional Hilbert space $\cH$ is unitary if $U^\dagger U=I$. A controlled‑$U$ applies $U$ to a target register conditional on a control qubit being $\ket{1}$.

\paragraph{Measurement and Born’s rule.}
A projective measurement is a collection of orthogonal projectors $\{P_i\}$ with $\sum_i P_i=I$. Measuring $\ket{\psi}$ returns outcome $i$ with probability $\bra{\psi}P_i\ket{\psi}$ and collapses the state to $P_i\ket{\psi}/\sqrt{\bra{\psi}P_i\ket{\psi}}$.

\paragraph{Quantum Fourier Transform.}
For an $r$‑qubit register, the QFT is the unitary
\[
\mathrm{QFT}_r:\;\ket{k} \;\mapsto\; \frac{1}{\sqrt{2^r}}\sum_{\ell=0}^{2^r-1} e^{2\pi i k\ell / 2^r}\ket{\ell},\qquad k\in\{0,\dots,2^r-1\}.
\]
Its inverse $\mathrm{QFT}_r^\dagger$ replaces $i$ by $-i$. In QBOOT, QFT is used inside amplitude estimation on a register of $T$ \emph{precision qubits}; we reserve $r$ here to avoid clashes and use $T$ exclusively for QAE precision in the main text.

\paragraph{Canonical quantum amplitude estimation.}

The canonical QAE algorithm uses two quantum registers. One register consists of $T$ qubits initialized in the state $\ket{0}$. We refer to it as \textit{measurement units}, and the initial state of this register is denoted by $\ket{0}^{\otimes^T}   = \underbrace{\ket{0}\otimes\cdots \otimes\ket{0}}_T$.
The joint system of the bootstrap sample register and bootstrap sample processor is initialized at the superposition 
$$\ket{\phi(z)} = \Ab \ket{\textbf{0}} = 2^{-1/2} ( e^{i (\tau_z-\pi/2)} \ket{\phi_+} +e^{-i (\tau_z-\pi/2)} \ket{\phi_-} ) .$$
The details of each step in the QAE algorithm using quantum phase estimation are summarized below.
    
\textit{Step 1.} Prepare the measurement units in the initial state of the uniform superposition.

We partially apply the Hadamard gate to all the $T$ qubits in the measurement units. Hence, the state of the joint system of measurement units, the bootstrap sample register, and the bootstrap sample processor becomes
\begin{eqnarray*}
   && (\frac{1}{\sqrt{2}}\ket{0}+\frac{1}{\sqrt{2}}\ket{1})^{\otimes^T}\otimes \ket{\phi(z)} \\
   &=& (\frac{1}{\sqrt{2}}\ket{0}+\frac{1}{\sqrt{2}}\ket{1})^{\otimes^T}  2^{-1/2} ( e^{\bfi (\tau_z-\pi/2)} \ket{\phi_+} +e^{-\bfi (\tau_z-\pi/2)} \ket{\phi_-} ).
\end{eqnarray*}


\textit{Step 2}. 
 For $k =1,...,
T$,  apply $2^{k-1}$ controlled-$\Qb$ operations to the $k$th joint system consisting of the $k$th qubit of the measurement units, the bootstrap sample register, and the bootstrap sample processor.

We first apply the controlled-$\Qb$ operation on
 the first qubit in the measurement units and all qubits of the bootstrap sample register and the bootstrap sample processor.  
 In particular, the joint system of the bootstrap sample register and the bootstrap sample processor is treated as the target register, and the qubit of the measurement units is treated as the control qubit.
Then, the state of the joint system becomes
\begin{eqnarray*}
&(\frac{1}{\sqrt{2}}\ket{0}+\frac{1}{\sqrt{2}}e^{\bfi 2 \tau_z}\ket{1})\otimes (\frac{1}{\sqrt{2}}\ket{0}+\frac{1}{\sqrt{2}}\ket{1})^{\otimes^{ T-1}}\otimes \frac{1}{\sqrt{2}}e^{\bfi( \tau_z-\pi/2)} \ket{\phi_+} +\\
&(\frac{1}{\sqrt{2}}\ket{0}+\frac{1}{\sqrt{2}}e^{-\bfi 2 \tau_z}\ket{1})\otimes (\frac{1}{\sqrt{2}}\ket{0}+\frac{1}{\sqrt{2}}\ket{1})^{\otimes^{ T-1}}\otimes \frac{1}{\sqrt{2}}e^{-\bfi( \tau_z-\pi/2)} \ket{\phi_-}.
\end{eqnarray*}

We continually apply the controlled-$\Gb_w^{2^{t-1}}$ operation on $t = 2,\cdots, T$th qubit in the measurement units in addition to the bootstrap sample register and the bootstrap sample processor. The state of the joint system becomes
\begin{equation}\label{eq:phase shift}
\frac{e^{\bfi( \tau_z-\pi/2)}}{2^{(T+1)/2}} \underset{t=1}{\overset{T}{\otimes}}
 (\ket{0}+e^{\bfi 2^t \tau_z}\ket{1})\otimes \ket{\phi_+}+\frac{e^{-\bfi( \tau_z-\pi/2)}}{2^{(T+1)/2}} \underset{t=1}{\overset{T}{\otimes}}
 (\ket{0}+e^{-\bfi 2^t \tau_z}\ket{1})\otimes \ket{\phi_-}.
\end{equation}
By elementary algebra, Eq.(\ref{eq:phase shift}) can be written as
\begin{equation}
    \frac{1}{2^{(T+1)/2}} \underset{k=0}{\overset{2^T-1}{\sum}}e^{\bfi \tau_z (2k+1) -\bfi\pi/2}
 \ket{k}\otimes \ket{\phi_+}+\frac{1}{2^{(T+1)/2}} \underset{k=0}{\overset{2^T-1}{\sum}}e^{-\bfi \tau_z (2k+1)-\bfi\pi/2}
 \ket{k}\otimes \ket{\phi_-}.
\end{equation}

\textit{Step 3}. Apply the inverse quantum Fourier transform to each qubit in the measurement units.

After we apply inverse quantum Fourier transform $QFT^\dagger$ on the measurement units, the state of the joint system becomes
\begin{equation}\label{eq:QFT}
    \ket{e_{\tau_z}}\otimes\frac{1}{\sqrt{2}}  \ket{\phi_+}+
 \ket{e_{-\tau_z}}\otimes\frac{1}{\sqrt{2}}  \ket{\phi_-},
\end{equation}
where $\ket{e_{\tau_z}} =  \frac{1}{2^{T}} \underset{k,l=0}{\overset{2^T-1}{\sum}} e^{\bfi\lt(\tau_z(2k+1)-\pi/2 - \frac{2\pi kl}{2^T} \rt) }\ket{l}$ and \\ $\ket{e_{-\tau_z}} =  \frac{1}{2^{T}} \underset{k,l=0}{\overset{2^T-1}{\sum}} e^{-\bfi\lt(\tau_z(2k+1) +\pi/2 + \frac{2\pi kl}{2^T}\rt) }\ket{l}$.

\textit{Step 4}. Measure the measurement units. The measurements of the measurement units are used to estimate $\tau_z$.
\section{Formal QBOOT Construction}
\label{app:qboot-formal}

\subsection{Encoding the empirical bootstrap measure}\label{app:encode}

Let $\bsX=(X_1,\ldots,X_n)$ denote the observed sample, and let $m=\lceil \log_2 n\rceil$.
We use \emph{index encoding} on $m$ qubits with computational basis
$\{\ket{0},\ldots,\ket{n-1}\}$ so that observation $X_i$ is associated with $\ket{i-1}$.
The empirical distribution $\hat{\PP}_n=n^{-1}\sum_{i=1}^n \delta_{X_i}$ is encoded by
\[
\ket{\hat{\PP}_n}\;=\;\frac{1}{\sqrt{n}}\sum_{i=1}^n \ket{i-1},
\]
so a projective measurement in the index basis returns index $i$ with probability $1/n$.
Placing $n$ such registers in parallel gives the exact quantum representation of the bootstrap product:
\begin{equation}
\label{eq:bootstrap-superposition}
\ket{\hat{\PP}_n^{\otimes n}}
\;=\;\bigotimes_{j=1}^n \ket{\hat{\PP}_n}
\;=\; \frac{1}{n^{n/2}} \sum_{\ib\in[n]^n} \ket{\ib},
\end{equation}
where $\ib=(i_1,\ldots,i_n)$ and $\ket{\ib}=\ket{i_1-1}\otimes\cdots\otimes\ket{i_n-1}$.
Each seed $\ib$ deterministically specifies the resampled dataset $\bsX^*(\ib)=(X_{i_1},\ldots,X_{i_n})$.

\subsubsection{Measurement–probability correspondence.}
Let $\cH_{\PP}=\mathrm{span}\{\ket{i-1}\}_{i=1}^n$ and
$\cP_n=\{\sum_{i=1}^n p_i\delta_{X_i}: p_i\ge 0,\ \sum_i p_i=1\}$.
For any $\ket{\psi}=\sum_{i=1}^n b_i\ket{i-1}\in\cH_{\PP}$, define the classical outcome map
\begin{equation}
\label{eq:meas-map}
\mathsf{Pr}(\ket{\psi}) \;=\; \sum_{i=1}^n |b_i|^2\,\delta_{X_i}\ \in\ \cP_n,
\end{equation}
which extends tensorially: $\mathsf{Pr}(\ket{\psi}\otimes\ket{\psi'})=\mathsf{Pr}(\ket{\psi})\otimes\mathsf{Pr}(\ket{\psi'})$.
Applying \eqref{eq:meas-map} to \eqref{eq:bootstrap-superposition} recovers the classical product measure $\hat{\PP}_n^{\otimes n}$ over resamples $\bsX^*(\ib)$.

\subsection{The indicator observable and its unitary realization}

Let $f:\Omega^n\to\RR$ be a statistic and $g(z,\bsX^*)=\mathbf{1}\{f(\bsX^*)\le z\}$.
Define the projector onto the “accept’’ subspace in the seed basis:
\[
\Mb_g(z)\;=\;\sum_{\ib:\, g(z,\bsX^*(\ib))=1}\ \ket{\ib}\bra{\ib}.
\]
By construction,
\[
\bra{\hat{\PP}_n^{\otimes n}} \Mb_g(z) \ket{\hat{\PP}_n^{\otimes n}} \;=\; H_{\mathrm{BOOT}}(z).
\]
To embed $g$ coherently, introduce a \emph{label qubit} and define a unitary $\Ub_g(z)$ acting as
\begin{equation}
\label{eq:oracle-Ug}
\Ub_g(z)\big(\ket{\ib}\ket{0}\big)\;=\;\ket{\ib}\,\ket{g\!\big(z,\bsX^*(\ib)\big)}.
\end{equation}
A block-diagonal construction in the orthonormal seed basis $\{\ket{\ib}\}_{\ib\in[n]^n}$ ensures the existence of $\Ub_g(z)$; a constructive realization follows by reversibly evaluating $f(\bsX^*(\ib))$, comparing to $z$, and uncomputing any workspace.

\subsection{State for amplitude estimation and eigenstructure}

Let $\Ub_{\hat{\PP}_n}$ be any state-preparation unitary with $\Ub_{\hat{\PP}_n}\ket{\mathbf{0}}_m=\ket{\hat{\PP}_n}$, and define
\[
\Ub_{\hat{\PP}_n^{\otimes n}}=\bigotimes_{j=1}^n \Ub_{\hat{\PP}_n},
\qquad
\Ab=\Ub_g(z)\,(\Ub_{\hat{\PP}_n^{\otimes n}}\otimes I).
\]
Acting on $\ket{\mathbf{0}}^{\otimes nm}\ket{0}$,
\[
\Ab\ket{\mathbf{0}} \;=\; 
\sqrt{H_{\mathrm{BOOT}}(z)}\,\ket{\phi_1} + \sqrt{1-H_{\mathrm{BOOT}}(z)}\,\ket{\phi_0},
\qquad \braket{\phi_1|\phi_0}=0,
\]
so the amplitude of $\ket{\phi_1}$ equals $\sqrt{H_{\mathrm{BOOT}}(z)}$.
Let $S_0=I-2\ket{\mathbf{0}}\bra{\mathbf{0}}$ (on all non-precision registers) and let $S_{\phi_1}$ flip the phase of $\ket{\phi_1}$.
The Grover iterate $\Qb=\Ab\, S_0\, \Ab^\dagger\, S_{\phi_1}$ has eigenvectors
\[
\ket{\phi_\pm}=\frac{\ket{\phi_1}\pm i\ket{\phi_0}}{\sqrt{2}},
\qquad
\Qb\ket{\phi_\pm}=e^{\pm i\,2\tau_z}\ket{\phi_\pm},
\]
with $\sin^2\tau_z=H_{\mathrm{BOOT}}(z)$.
Applying QAE with $T$ precision qubits to $\Qb$ yields the estimator
\[
H_{\mathrm{QBOOT}}(z)=\sin^2\!\left(\frac{\pi Y}{2^T}\right),
\qquad Y\in\{0,\ldots,2^T-1\}.
\]


\section{Implementation Details}
\label{app:implementation}

\paragraph{State preparation (bootstrap encoding unit).}
With \emph{index encoding}, $\ket{\psi_{X_i}}=\ket{i-1}$ on $m=\lceil\log_2 n\rceil$ qubits and $\ket{\hat{\PP}_n}$ is prepared by a uniform superposition, followed by a classical table that maps indices to data values when needed. 
When downstream arithmetic must act on the values themselves, one may use \emph{value encodings} (e.g., basis, angle, or amplitude encodings) in a data register linked to the index register.
Equation~\eqref{eq:bootstrap-superposition} is then prepared by $n$ parallel copies of $\Ub_{\hat{\PP}_n}$.

\begin{figure}[!htb]
  \centering
  \includegraphics[width=0.95\textwidth]{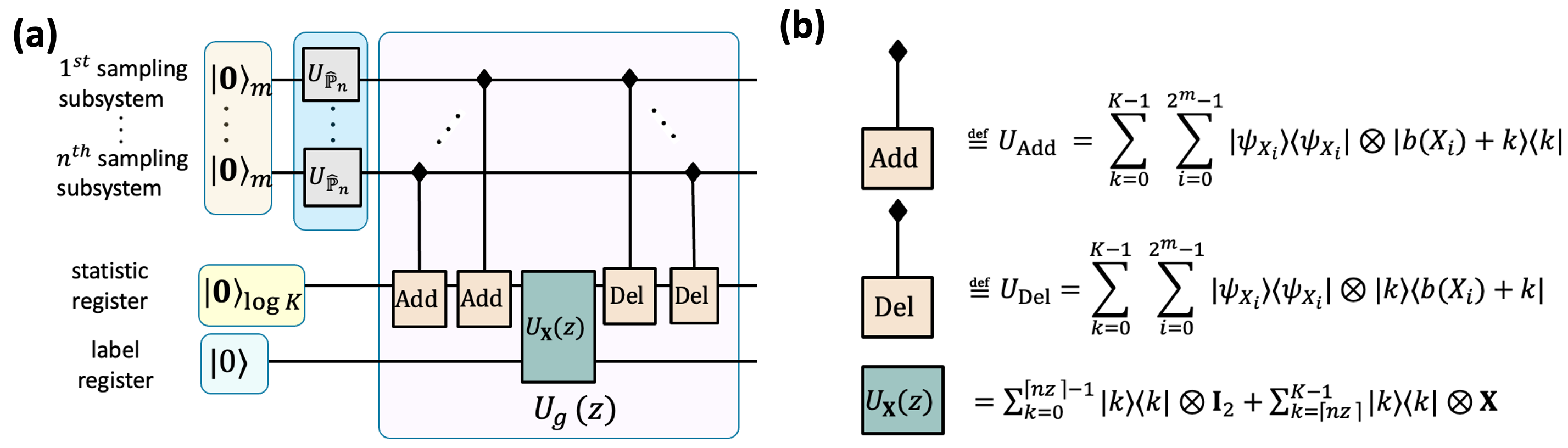}
  \caption{Example realisation of $\Ub_g(z)$ for the sample‑mean indicator $g(z,\bsX^*)=\mathbf{1}\{\bar X^*\le z\}$. The circuit accumulates $\sum_j X_j^*$ in a statistic register, compares to $z$, flips a label qubit, and uncomputes the arithmetic.}
  \label{fig:QC_mean_exmp}
\end{figure}

\paragraph{Statistic evaluation (estimator unit).}
{A practical $\Ub_g(z)$ mirrors the classical computation of $g$ with reversible arithmetic:
accumulate (in a statistic register) the quantity required by $f$, compare to $z$, write the comparison result to the label qubit, and uncompute the accumulator.
For the sample mean, the circuit implements a reversible adder to compute $\sum_{j=1}^n X_j^*$, a comparator with threshold $n z$, a controlled $X$ on the label qubit, and then the inverse adder to return all ancillae to $\ket{0}$, as shown in Figure \ref{fig:QC_mean_exmp}.
This realisation typically yields quantum gate complexity $Q_g$ comparable to the classical cost $C_g$; specific $g$ may admit quantum subroutines with $Q_g\ll C_g$ (e.g., Grover‑style routines for order statistics).}

\begin{table}
\caption{Comparison between $Q_g$ and $C_g$ on common statistics calculation \\ and machine learning model training.}
\label{tab:Q_g_vs_C_g}
\begin{tabular}{ccc}
\toprule
 & \textbf{$Q_g$} & \textbf{$C_g$} \\
\midrule
Rank statistics for $n$-size unsorted sample \cite{grover1996fast,nayak1999quantum} & $\cO(\sqrt{n})$ & $\cO(n)$ \\
\hline
Bayesian Inference\citep{low2014quantum} & $\cO(\epsilon^{-1} )$ & $\cO(\epsilon^{-2} )$ \\
\hline
PCA\citep{lloyd2014quantum} & $\cO(\log(n) )$ & $\cO(n )$ \\
\hline
Linear Regression\citep{schuld2016prediction} & $\cO(\log(n) )$ & $\cO(n )$ \\
\hline
Kernel Regression\citep{PhysRevLett.122.040504} & $\cO(\log(n) )$ & $\cO(n^3 )$ \\
\hline
SVM \citep{rebentrost2014quantum} & $\cO(\log(n) )$ & $\cO(n^3 )$ \\
\bottomrule
\end{tabular}
\end{table}

\paragraph{Amplitude estimation (estimation unit).}
Canonical QAE with $T$ precision qubits uses $\mathcal{O}(2^T)$ controlled powers of $\Qb$ and one inverse $\mathrm{QFT}^\dagger$.
On noisy devices, iterative or likelihood‑based amplitude‑estimation variants can reduce depth at the cost of constant factors in the number of oracle calls.
In either case, repeating QAE a small number of times and taking the median of the resulting $H_{\mathrm{QBOOT}}(z)$ values is effective at suppressing occasional large errors caused by the oscillatory tails of the phase‑estimation distribution.

\section{Proof Sketches Underpinning the Main Results}

Let $H_{\mathrm{BOOT}}(z)=\sin^2\tau_z$.

\begin{figure}[!htb]
  \centering
  \includegraphics[width=0.85\textwidth]{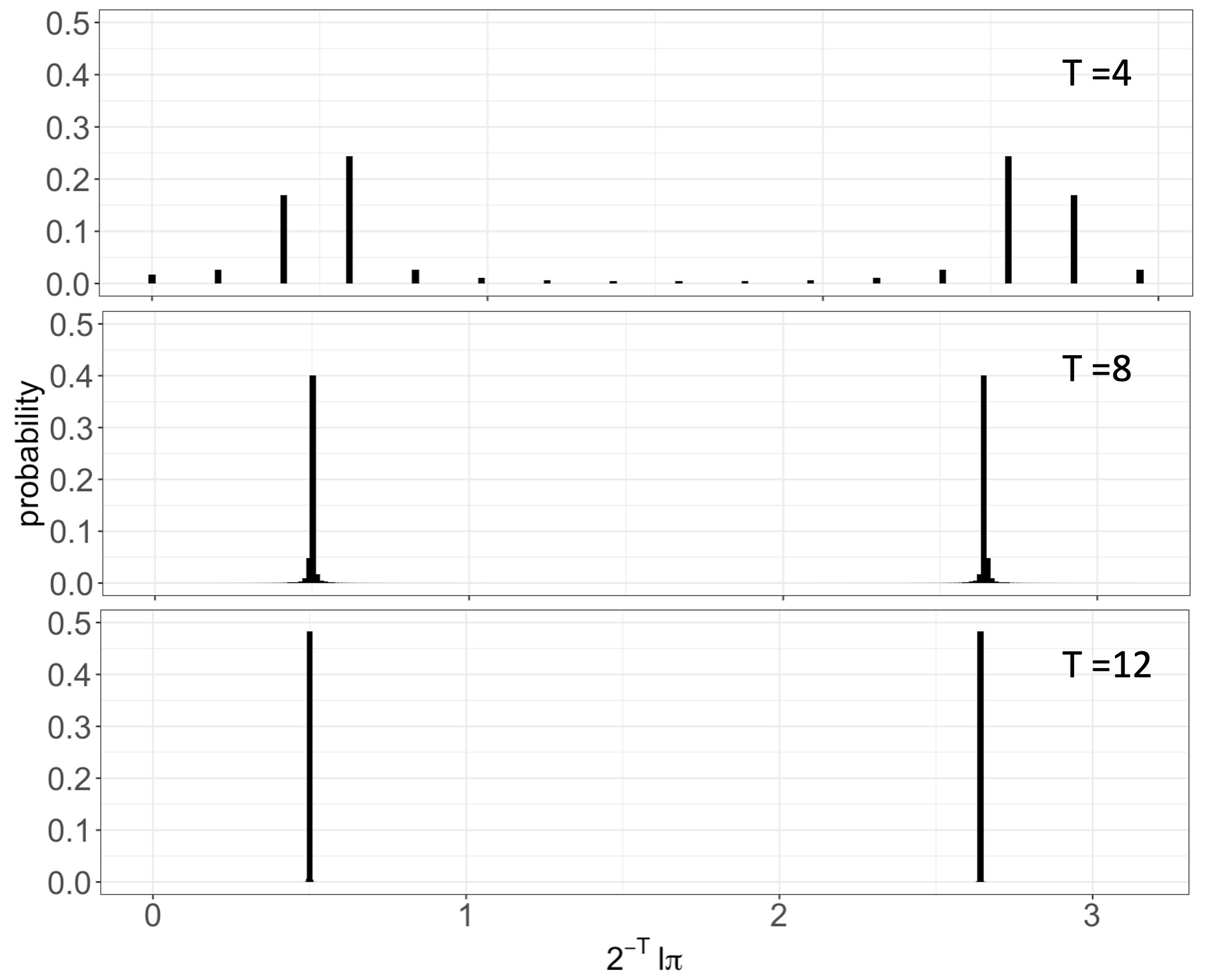}
  \caption{Probability distribution of QAE outcomes $\Pr(l_z=l)$ for $\tau_z=0.5$ and $T\in\{4,8,12\}$, illustrating concentration near $2^{-T}l\pi\approx \tau_z$ that drives the $\mathcal{O}_p(2^{-T})$ error bound.}
  \label{fig:probs_exmp}
\end{figure}

\paragraph{Measurement outcome distribution.}
A standard calculation for phase estimation applied to $\Qb$ gives, for $l\in\{0,\ldots,2^T-1\}$,
\begin{equation}
\label{eq:pe-distribution}
\Pr(l_z=l)\;=\;
\frac{\sin^2(2^T\tau_z)}{2^{2T+1}}
\left[
\csc^2\!\left(\tau_z - \frac{l\pi}{2^T}\right)
+
\csc^2\!\left(\tau_z + \frac{l\pi}{2^T}\right)
\right].
\end{equation}
The mass concentrates near $l$ such that $2^{-T}l\pi\approx \tau_z$ or $\pi-\tau_z$.
The example of probability distribution of QAE outcomes $\Pr(l_z=l)$ for $\tau_z=0.5$ and $T\in\{4,8,12\}$ is shown in Figure \ref{fig:probs_exmp}.
\paragraph{Quantum error bound (Theorem~\ref{thm:qerror}).}
Using \eqref{eq:pe-distribution}, one bounds the probability that $Y$ falls outside the nearest integers to $2^T\tau_z/\pi$; mapping $Y\mapsto \sin^2(\pi Y/2^T)$ then yields
$K(H_{\mathrm{QBOOT}},H_{\mathrm{BOOT}})=\mathcal{O}_p(2^{-T})$.

\paragraph{Bias and MSE (Theorem~\ref{thm:biasmse}).}
Taking expectations of $H_{\mathrm{QBOOT}}(z)-H_{\mathrm{BOOT}}(z)$ and its square with respect to \eqref{eq:pe-distribution} gives the exact expressions
\begin{align}
\mathbb{E}[H_{\mathrm{QBOOT}}(z)] - H_{\mathrm{BOOT}}(z)
&= 2^{-2T}\sin^2(2^T\tau_z)\sum_{l=0}^{2^T-1}\frac{\sin(2^{-T}l\pi+\tau_z)}{\sin(2^{-T}l\pi-\tau_z)}, \nonumber\\
\mathbb{E}\!\left[(H_{\mathrm{QBOOT}}(z)-H_{\mathrm{BOOT}}(z))^2\right]
&= 2^{-2T}\sin^2(2^T\tau_z)\sum_{l=0}^{2^T-1}\sin^2(2^{-T}l\pi+\tau_z). \nonumber
\end{align}
Approximating the sums by integrals shows both the bias and the MSE are $\mathcal{O}(2^{-T})$ uniformly in $z$ (Corollary~\ref{cor:asymptotic}), and explains the observed oscillations as functions of $H_{\mathrm{BOOT}}(z)$.

{
\section{Proof of the Median-of-$M$ Theorem}
\label{app:proof-median}

\paragraph{Single-run error bound(Lemma~\ref{prop:single-run}).}
Fix $z\in\mathbb{R}$ and write $\tau_z=\arcsin\sqrt{H_{\mathrm{BOOT}}(z)}\in[0,\pi/2]$.
Let $l^*=\lfloor 2^T\tau_z/\pi\rfloor$ and $\phi=2^T\tau_z/\pi-l^*\in[0,1)$.
Because $\sin^2(\pi l/2^T)=\sin^2(\pi(2^T-l)/2^T)$, 
$\{|H_{\mathrm{QBOOT}}(z)-H_{\mathrm{BOOT}}(z)|\le\varepsilon_T\}$
contains four outcomes:
\[
  \mathcal{S}_z \;=\; \{l^*,\;l^*+1,\;2^T-l^*-1,\;2^T-l^*\}.
\]
We consider
$\mathcal{S}^+=\{l^*,l^*+1\}$  and $\mathcal{S}^-=\{2^T-l^*-1,2^T-l^*\}$ .

For $\mathcal{S}^+$,
using $\csc^2(x)\ge 1/x^2$ (since $\sin x\le x$) and Theorem \ref{thm:exact_dist}, we have
\begin{equation}
  \Pr(Y_z\in\mathcal{S}^+)
  \;\ge\; \frac{\sin^2(\phi\pi)}{2^{2T+1}}
  \left[\frac{2^{2T}}{\phi^2\pi^2}+\frac{2^{2T}}{(1-\phi)^2\pi^2}\right]
  \;=\; \frac{\sin^2(\phi\pi)}{2\pi^2}
  \left[\frac{1}{\phi^2}+\frac{1}{(1-\phi)^2}\right].
  \label{eq:pos-pair}
\end{equation}
 
Similarly,
\begin{equation}
  \Pr(Y_z\in\mathcal{S}^-)
  \;\ge\; \frac{\sin^2(\phi\pi)}{2\pi^2}
  \left[\frac{1}{(1-\phi)^2}+\frac{1}{\phi^2}\right].
  \label{eq:alias-pair}
\end{equation}
 
Adding \eqref{eq:pos-pair} and \eqref{eq:alias-pair}:
\[
  q_z \;\ge\; \frac{\sin^2(\phi\pi)}{\pi^2}
  \left[\frac{1}{\phi^2}+\frac{1}{(1-\phi)^2}\right]
  \;=:\; h(\phi).
\]
Evaluating at the minimum of $h(\phi)$, we have 
$h(\phi) \geq h(1/2)=\sin^2(\pi/2)\cdot[4+4]/\pi^2=8/\pi^2$.
Therefore
\begin{equation}
  q_z \;\ge\; \frac{8}{\pi^2} \;>\; \frac{1}{2},
  \label{eq:qz-lower}
\end{equation}
with equality when $\phi=1/2$, i.e., when $\tau_z$ falls exactly halfway
between two adjacent grid points.
 
 \paragraph{Median-of-$M$ error bound(Theorem~\ref{thm:median-M}).}
Let $X_m=\mathbf{1}\{|H_{\mathrm{QBOOT}}^{(m)}(z)-H_{\mathrm{BOOT}}(z)|>\varepsilon_T\}$
be i.i.d.\ Bernoulli with success parameter $1-q_z\le 1-8/\pi^2<1/2$.
Set $S=\sum_{m=1}^M X_m\sim\mathrm{Binomial}(M,1-q_z)$.
The median fails only if $S\ge\lceil M/2\rceil$.
For any $t>0$:
\[
  \Pr(S\ge M/2)
  \;\le\; e^{-tM/2}\,\mathbb{E}[e^{tS}]
  \;=\; e^{-tM/2}\,(1-q_z e^t + q_z)^M.
\]
Minimizing over $t>0$ by setting the derivative to zero gives
$e^{t^*}=q_z/1-q_z$, and substituting:
\begin{equation}
  \Pr(S\ge M/2)
  \;\le\;
  \left(2\sqrt{1-q_z\,q_z}\right)^M
  \;=\;
  \bigl(4(1-q_z) q_z\bigr)^{M/2}.
  \label{eq:chernoff-binom}
\end{equation}
Since 
\[
  4(1-q_z) q_z\;=\; 4\!\left(1-\tfrac{8}{\pi^2}\right)\tfrac{8}{\pi^2}
  \;\approx\; 0.614.
\]
 
Combining the results, we have
\[
  \Pr\!\left(|H_{\mathrm{QBOOT}}^{(M)}(z)-H_{\mathrm{BOOT}}(z)|>\varepsilon_T\right)
  \;\le\; \rho^{M/2}.
\]
Setting $\rho^{M/2}=\delta$, we have 
$M\ge 2\log(1/\delta)/\log(1/\rho_0)\approx4.11\log(1/\delta)$.
 }
 \newpage
\section{Figures}

\end{document}